\documentclass[10pt,twocolumn,twoside,aps,pra,nopacs,superscriptaddress,a4paper,nofootinbib]{revtex4-1}
\usepackage{graphicx}
\usepackage{amsmath,amssymb}
\usepackage{color}
\usepackage{bbm,subfigure}
\usepackage[colorlinks=true,urlcolor=blue]{hyperref}
\usepackage{theorem}
\usepackage{latexsym}

\newlength{\blank}
\newcommand{\ket}[1]{|#1\rangle}

\mathchardef\ordinarycolon\mathcode`\:
\mathcode`\:=\string"8000
\def\vcentcolon{\mathrel{\mathop\ordinarycolon}}
\begingroup \catcode`\:=\active
  \lowercase{\endgroup
  \let :\vcentcolon
  }

\newcommand{\nc}{\newcommand}
\nc{\rnc}{\renewcommand}
\nc{\beq}{\begin{equation}}
\nc{\eeq}{{\end{equation}}}
\nc{\beqa}{\begin{eqnarray}}
\nc{\eeqa}{\end{eqnarray}}
\nc{\lbar}[1]{\overline{#1}}
\nc{\ketbra}[2]{|#1\rangle\!\langle#2|}
\nc{\proj}[1]{| #1\rangle\!\langle #1 |}
\nc{\avg}[1]{\langle#1\rangle}
\nc{\Rank}{\operatorname{Rank}}
\nc{\smfrac}[2]{\mbox{$\frac{#1}{#2}$}}
\nc{\tr}{\operatorname{Tr}}
\nc{\ox}{\otimes}
\nc{\dg}{\dagger}
\nc{\dn}{\downarrow}
\nc{\cA}{\mathcal{A}}
\nc{\cB}{\mathcal{B}}
\nc{\cC}{\mathcal{C}}
\nc{\cD}{\mathcal{D}}
\nc{\cE}{\mathcal{E}}
\nc{\cF}{\mathcal{F}}
\nc{\cG}{\mathcal{G}}
\nc{\cH}{\mathcal{H}}
\nc{\cI}{\mathcal{I}}
\nc{\cJ}{\mathcal{J}}
\nc{\cK}{\mathcal{K}}
\nc{\cL}{\mathcal{L}}
\nc{\cM}{\mathcal{M}}
\nc{\cN}{\mathcal{N}}
\nc{\cO}{\mathcal{O}}
\nc{\cP}{\mathcal{P}}
\nc{\cR}{\mathcal{R}}
\nc{\cS}{\mathcal{S}}
\nc{\cT}{\mathcal{T}}
\nc{\cX}{\mathcal{X}}
\nc{\cZ}{\mathcal{Z}}
\nc{\csupp}{{\operatorname{csupp}}}
\nc{\qsupp}{{\operatorname{qsupp}}}
\nc{\var}{\operatorname{var}}
\nc{\rar}{\rightarrow}
\nc{\lrar}{\longrightarrow}
\nc{\polylog}{\operatorname{polylog}}
\nc{\1}{{\openone}}
\nc{\id}{{\operatorname{id}}}

\nc{\RR}{{{\mathbb R}}}
\nc{\CC}{{{\mathbb C}}}
\nc{\FF}{{{\mathbb F}}}
\nc{\NN}{{{\mathbb N}}}
\nc{\ZZ}{{{\mathbb Z}}}
\nc{\PP}{{{\mathbb P}}}
\nc{\QQ}{{{\mathbb Q}}}
\nc{\UU}{{{\mathbb U}}}
\nc{\EE}{{{\mathbb E}}}

\nc{\qed}{{$\hfill\Box$}}

\begin{document}

\title{Quantum Godwin's Law}

\author{Michalis Skotiniotis${}^{+,}$}
%\email{michail.skoteiniotis@uab.cat}
%\affiliation{Departament de F\'{\i}sica, Grup d'Informaci\'{o} Qu\`{a}ntica, Universitat Aut\`{o}noma de Barcelona, 08193 Bellaterra (Barcelona), Spain}
\affiliation{UAB, ${}^+$IAF \&{} ${}^-$ICREA}

\author{Andreas Winter${}^{-,}$}
%\email{andreas.winter@uab.cat}
%\affiliation{ICREA -- Instituci\'{o} Catalana de Recerca i Estudis Avan\c{c}ats, Pg.~Llu\'{\i}s Companys, 23, 08010 Barcelona, Spain}
%\affiliation{Departament de F\'{\i}sica, Grup d'Informaci\'{o} Qu\`{a}ntica, Universitat Aut\`{o}noma de Barcelona, 08193 Bellaterra (Barcelona), Spain}
\affiliation{UAB, ${}^+$IAF \&{} ${}^-$ICREA}

\date{1 April 2020}

\begin{abstract}
Godwin's law, i.e. the empirical observation that as an online discussion 
grows in time, the probability of a comparison with Nazis or Hitler 
quickly approaches unity, is one of the best-documented facts of the internet. 
Anticipating the quantum internet, here we show under reasonable model 
assumptions a polynomial quantum speedup of Godwin's law. Concretely, 
in quantum discussions, Hitler will be mentioned on average quadratically 
earlier, and we conjecture that under specific network topologies, 
even cubic speedups are possible. We also show that the speedup cannot
be more than exponential, unless the polynomial hierarchy collapses to 
a certain finite level. 

We report on numerical experiments to simulate the appearance of 
the quantum Godwin law in future quantum internets; the most amazing 
finding of our studies is that -- unlike quantum computational speedups --
the quantum Godwin effect is not only robust against noise, but actually 
enhanced by decoherence. We have as yet no theoretical explanation, nor 
a good application, for this astonishing behaviour, which we dub 
\emph{quantum hyperpiesia}. 
%a term we rather hope doesn't stick. 
\end{abstract}

\maketitle

%%%%%%%%%%%%%%%%%%%%%%%%%%%%%%%%%%%%%%%%%%%%%%%%%%%%%%%%%%%%%%%%%%%

\noindent
\textbf{Introduction.}
%\label{sec:intro}
The quantum internet is a mystical entity, beckoning ever 
larger number of researchers and stake-holders that are looking for the 
holy grail of a thumping real-world application, if not eternal funding.
Its development, however, requires hard work and considerable 
ingenuity, with many obstacles remaining
\cite{Kimble:QI,Wehner:QI,DahlbergWehner,Gyongyosi-Imre-1,Gyongyosi-Imre-1.5,Gyongyosi-Imre-2,Gyongyosi-Imre-3,Gyongyosi-Imre-4,Castelvecchi}.

%\begin{quote}
\emph{In what can be identified as typical expert behaviour, the scientists 
focus largely on the technical aspects of realising the quantum 
internet, without pausing at all to consider \underline{why} we should do it,
and even less \underline{who} the eventual retarded users of their splendid 
quantum internet might be. 
We think it is time to confront these questions before another couple
billion dollars are poured down that drain. 
Only a bunch of nerds like the majority of the readers of the 
present paper can think that such an expense of resources is 
justified by the eventual possibility of Alice, Bob and some 
of their boffin friends being able to exchange provably secret 
messages. Which, let's face it, no-one really cares about anyway.}
%\end{quote}

Before we continue, we hasten to point out that the previous 
paragraph only serves as a typical example of online interactions, 
which are characterised by sweeping generalisations and apodictic 
proclamations rather than differentiated analysis, and, 
crucially, crude \emph{argumenta ad hominem} against the perceived 
opponent. It is here that apparently invariably, someone or something 
will be compared to Adolf Hitler.
The latter propensity, manifest in countless flame wars and 
blog rants, has been recognised as a fundamental fact of online life,
and borders on a law of nature \cite{onion}, which is indeed how 
Mike Godwin has conceived of it \cite{Godwin}. The empirical evidence 
of what is commonly identified as Godwin's law,
\begin{quote}
  \emph{As an online discussion grows longer, the probability of a 
  comparison involving Nazis or Hitler approaches one,}
\end{quote}
is enormous and ever mounting, and has been shown to determine discourse 
even long before the advent of the internet \cite{Orwell:Hitler}. 
Recent evidence suggests that it even governs monologues \cite{Woody}. 
Quantitative investigations, however, into the time it 
takes to reach that point are scarce. 

\medskip
In the present paper we not only fill this much-needed gap, by 
providing a toy model network in which we can exactly quantify 
the expected time until a Hitler comparison occurs, but we show that 
quantum internets offer at least a quadratic speedup. 
We can explicitly calculate the respective expected times for
a family of toy model networks, and we present assorted conjectures 
for general networks. 
The most important discovery, so far only numerically supported, 
is the astonishing resilience of the quantum Godwin speedup to, 
and even enhancement by decoherence, a novel quantum effect in need 
of a theoretical explanation and a constructive application. 

The rest of the paper consists of more or less tedious technicalities, 
which the impatient expert reader may want to skip, to check directly if 
their own work has been cited in the references~\cite{unknown}.

\bigskip\noindent
\textbf{Methodological preamble.}
%\label{sec:bla}
Godwin's law concerns the semantic universe demarcated by the words
Hitler, Nazi, Fascist, Francoist, and others appropriated from 
the military sphere (such as \emph{panzer}, \emph{blitzkrieg}, \emph{schadenfreude}, 
etc), but also concentration camp, Holocaust and Nuremberg process, 
whose meaning and connotations have a diverse and much-studied history, 
the subtle and not-so-subtle distinctions between them having been the subject of 
considerable 
%historical, philosophical and sociological 
erudition. 
However, as they are roundly irrelevant in online discussions (cf. \cite{jokes}), 
for the present paper we shall consider them all the same.
%we shall treat them as all the same. 

\begin{figure*}[ht]
  \includegraphics[width=3cm]{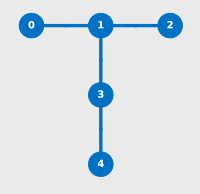}
  \includegraphics[width=3cm]{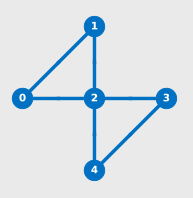}
  \includegraphics[width=8cm]{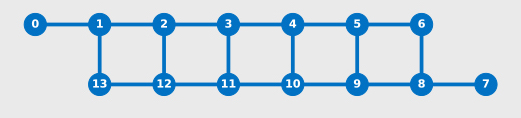}
  \includegraphics[width=4cm]{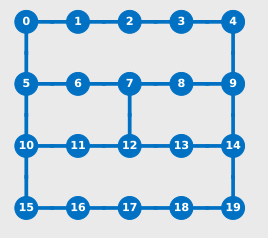}
  \includegraphics[width=4cm]{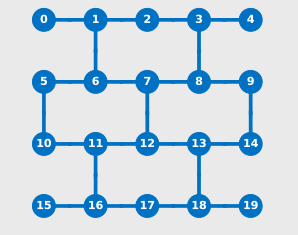}
  \includegraphics[width=7cm]{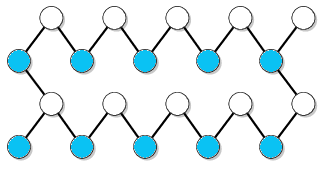}
  \includegraphics[width=7cm]{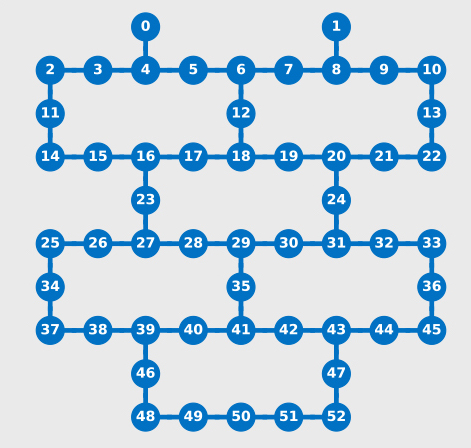}
  \includegraphics[width=7cm]{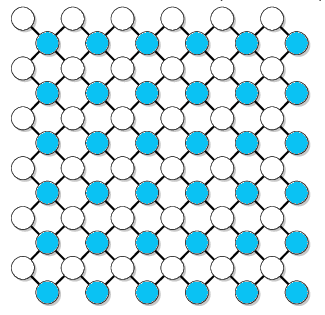}
  \caption{The first elements in the sequence of network topologies for which 
           the classical Godwin law is quantifiable.}
  \label{fig:fig}
\end{figure*}

Note furthermore that for an exact scientific discourse, 
Godwin's law is stated in terms too vague to be useful. This is not because 
it is too general to be true, but rather that it is true for trivial 
reasons that have not much to do with the nature of online discussions
or Nazis. Imagine a hypothetical Toto's law,
%Bart Simpson's law,
\begin{quote}
  \emph{As a discussion grows longer, the probability of 
  your dog being mentioned approaches one.}
\end{quote}
This holds true regardless of whether one has a dog, because without qualifications, 
``as a discussion grows longer'' is an asymptotic statement. If a conversation 
is a stochastic process, only the mildest assumptions of ergodicity or 
mixing need to be made to have any sequence of words appear with probability 
one over unlimited times of observation. 

It can be surmised that Godwin's intention was not to claim the obvious, 
that by the law of large numbers any concept is eventually mentioned, 
even admitting that the Nazi comparison is perceived as some sort of 
attractor of online discussions. 
All this shows is simply that Godwin's law needs to be augmented by a 
quantitative statement as to how soon someone will be compared to Hitler. 

We have been able to identify a sequence of model network topologies 
(``internets''), of which the first few members are shown in Fig. \ref{fig:fig}. 
By a remarkable coincidence, they replicate precisely the connectivity of
the first primitive superconducting universal quantum processors that have 
been realised in recent years in various labs \cite{IBM,RigettiGoogle}. 

The networks $G_n$, one for each integer $n$, are defined recursively by a 
standard VCP (very complex process); 
$G_n$ has a number of nodes that is a quite complicated function of $n$, 
but we know that it is monotonically non-decreasing and $\sim e^{n}$ for 
large $n$. The precise definition and discussion of its properties can
be found in Appendix \ref{app:networks}. The idea is that the network 
represents the internet, where an online discussion
takes place. Each node represents a generic internet user, and the
connectivities represent to which other users' contributions 
they react.

As can be seen from the definition, the network $G_n$ is such that the 
expected time until a Nazi comparison is 
\begin{equation}
  t_n = t(G_n) = n\left(\sum_{k=1}^n \frac{1}{k!}\right) \sim n(e-1). 
\end{equation}
This sets the classical baseline against which we can compare the
performance of the quantum internet.

\bigskip\noindent
\textbf{Quantum Godwin's law.}
%\label{sec:main}
Now that we have built our straw man, we can attack it. Concretely, 
we propose to consider each classical graph as a quantum network, with 
one user for each node, capable of quantum transformations and accessing 
both private and a common quantum memory; the latter we call the conversation,
and the graph structure determines how each user accesses the quantum 
conversation. The correspondence with the classical network and its dynamics is 
described by the principles of half and quarter quantisation. The reader
not familiar with these methods is invited to consult Appendix \ref{app:half},
which, while not exactly answering their questions, should keep them occupied 
for a while. General background on quantum information theory can be found 
in many an excellent textbook, such as \cite{NielsenChuang}, alternatively 
\cite{nutter} (see however \cite{jesus}).

With the basic notions out of the way, we start with 
a simple explicit calculation of the expected time 
$\tau_n=\tau(G_n)$ of quantum evolution of the network until Hitler is mentioned 
in $G_n$. It follows essentially from the definitions that
\begin{equation}
  \tau_n \sim \sqrt{2n}, 
\end{equation}
quadratically faster than the classical Nazi comparison. 

More generally (Appendix \ref{app:yetanother} or whatever), it can be shown 
that among neutral networks, 
if $G$ has chromatic pagenumber $p$, it holds $\tau(G) \leq O(2^{\sqrt{p}})$, 
while $t(G) \geq \sqrt{2}^{\sqrt{p}}$ for almost all such graps, 
by Euler's classic result. 

We conjecture, in line with the widely accepted hypothesis that Euler's 
bound is not optimal, that cubic separations are possible \cite{Piffle}.

\bigskip\noindent
\textbf{Upper and lower bounds on quantum Godwin.}
%\label{sec:PH}
We do not know what the largest separation between $t(G)$ and $\tau(G)$ is.
In fact, already the behaviour of $t(G)$ is poorly understood, since it 
is popularly believed to be uncomputable. The argument rests on the fundamentally
self-referential nature of the internet and its laws \cite{internet-laws}, 
which means that Godwin's law holds even among people who are aware of 
it, and in contexts where it has been explicitly flagged. The only possible 
explanation seems to be that in generic networks, Hitler is mentioned
sooner than any algorithm could have predicted, even taking into account 
Godwin's law (cf. Hofstadter's law). 
It follows that in the natural, lexicographic enumeration of graphs, 
the first index $|G|$ of a random graph taking time $>t$ is at least as large 
as $\mathsf{BB}(t)$, the $t$-th \emph{busy beaver number}, 
%\footnote{Are we still doing phrasing?}
with probability approaching $1$. 

Interestingly, by a unique piece of mathematical magic, we can 
show that if there were infinitely many graphs for which 
$\tau(G) < \ln t(G)$, the polynomial hierarchy would collapse to a finite 
level $L$, leading to a world-wide vendetta against beavers and 
an unprecedented shortage of toilet paper.
%Maybe we can adapt this here to include the beaver number... 
It is hard though to predict which level; 
the best estimate we have for $L$ is that it cannot be larger than 
$\mathfrak{p}$, the largest prime among the
wild numbers, as defined by A. Millechamps de Beauregard in 1823 \cite{Swift}.
%
%...the Collatz number $\mathfrak{C}$, defined as the smallest integer for which 
%the Collatz (aka Ulam, aka Kakutani, aka Hasse, aka Thwaites, aka $(3n+1)$-rule)
%sequence does not terminate in $1$. 

Let us be honest, however: no-one wants to see the details of this, so if 
you must, go and bother Scott Aaronson on his blog about it, and the
many intelligent people commenting there \cite{GreatScott}.

\bigskip\noindent
\textbf{Experiment and simulation.}
%\label{sec:exp}
Having not much else left to do, we decided to test the quantum Godwin
law both in experiments and simulations. Fortuitously, some of the 
networks $G_n$ in Fig.~\ref{fig:fig} have been built by dedicated research 
teams around the world.  Unfortunately, as most of the architectures are not publicly 
accessible -- and our funding agencies staunchly refused to pay the exorbitant fees
needed for their use --, we resorted to bribery, extortion, and name-calling.  
None of this worked,\footnote{Don't talk to us about not publishing negative results!} 
so eventually we had to make do with 
counterfactual quantum computation~\cite{counterf}, supplemented by 
%weeks-long 
arduous numerical simulation using counterfactual classical computations,
which were executed by several thousand anonymous Oompa Loompas (AOL).

After initial calibration of the quantum network parameters, the experiment, as well as the simulation, 
where run under noiseless, as well as noisy environments. The latter involve purely classical, 
or quantum noise models with or without memory, as well as hybrid quantum-classical noise models, 
too.
The implementation of such models posed the biggest challenge for the AOLs,
%Oompa Loompas, 
and was achieved deploying classical and quantum machine learning 
methods~\cite{SuttonBarto,Biamonte} to calculate the stochastic nature of the 
network as well as the associated probabilities. 
The resulting technique, which we call 
Asynchronous Randomisation and Convolutional Hitlerianism Error Ramblings
(``Archer''), is of interest in itself and we believe that 
subsequent work on quantum internet speedups may profoundly benefit from its use.

\begin{figure*}[ht]
\centering
\subfigure[]{\includegraphics[width=6cm]{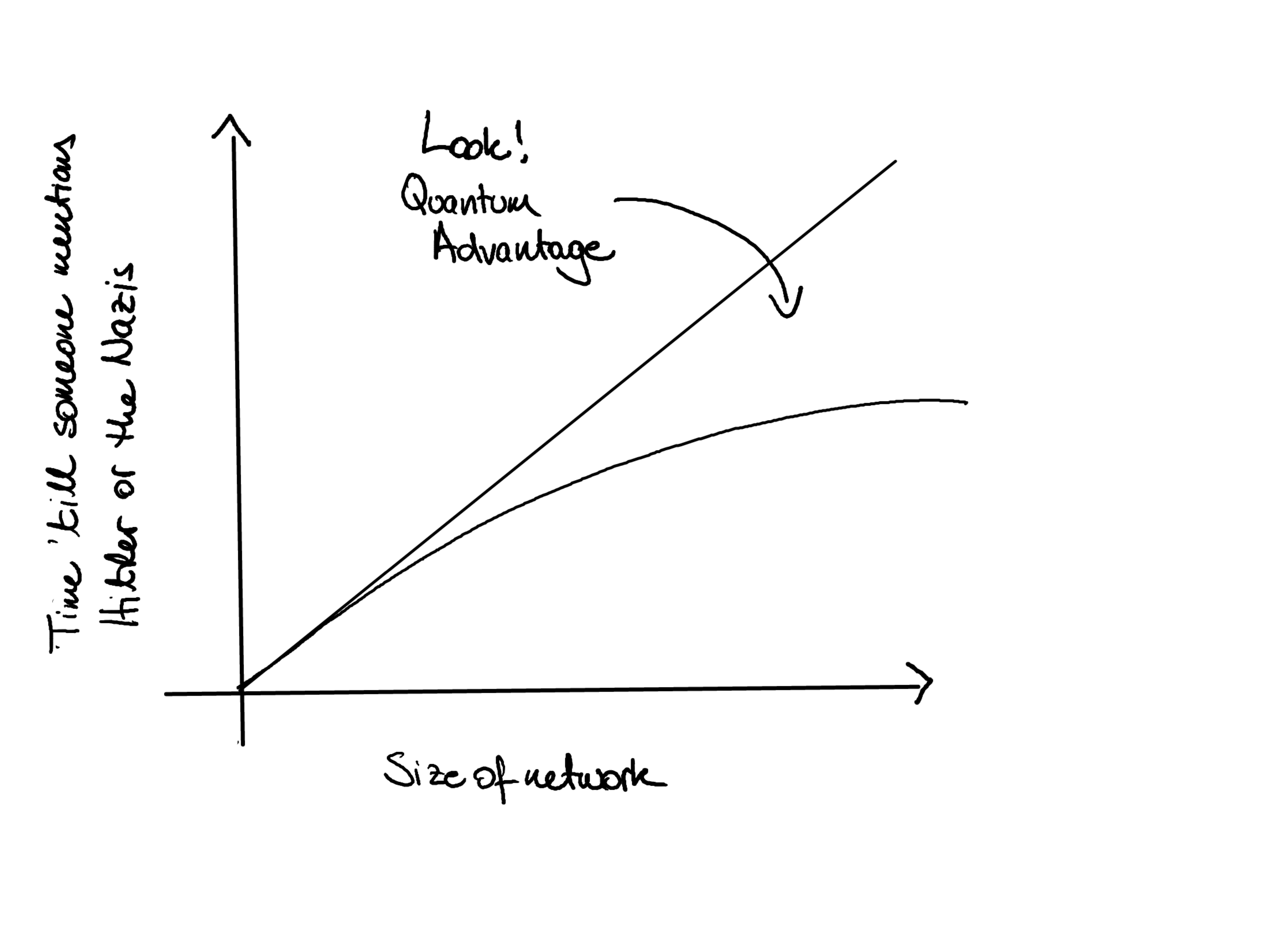}}
\qquad
\subfigure[]{\includegraphics[width=5.7cm]{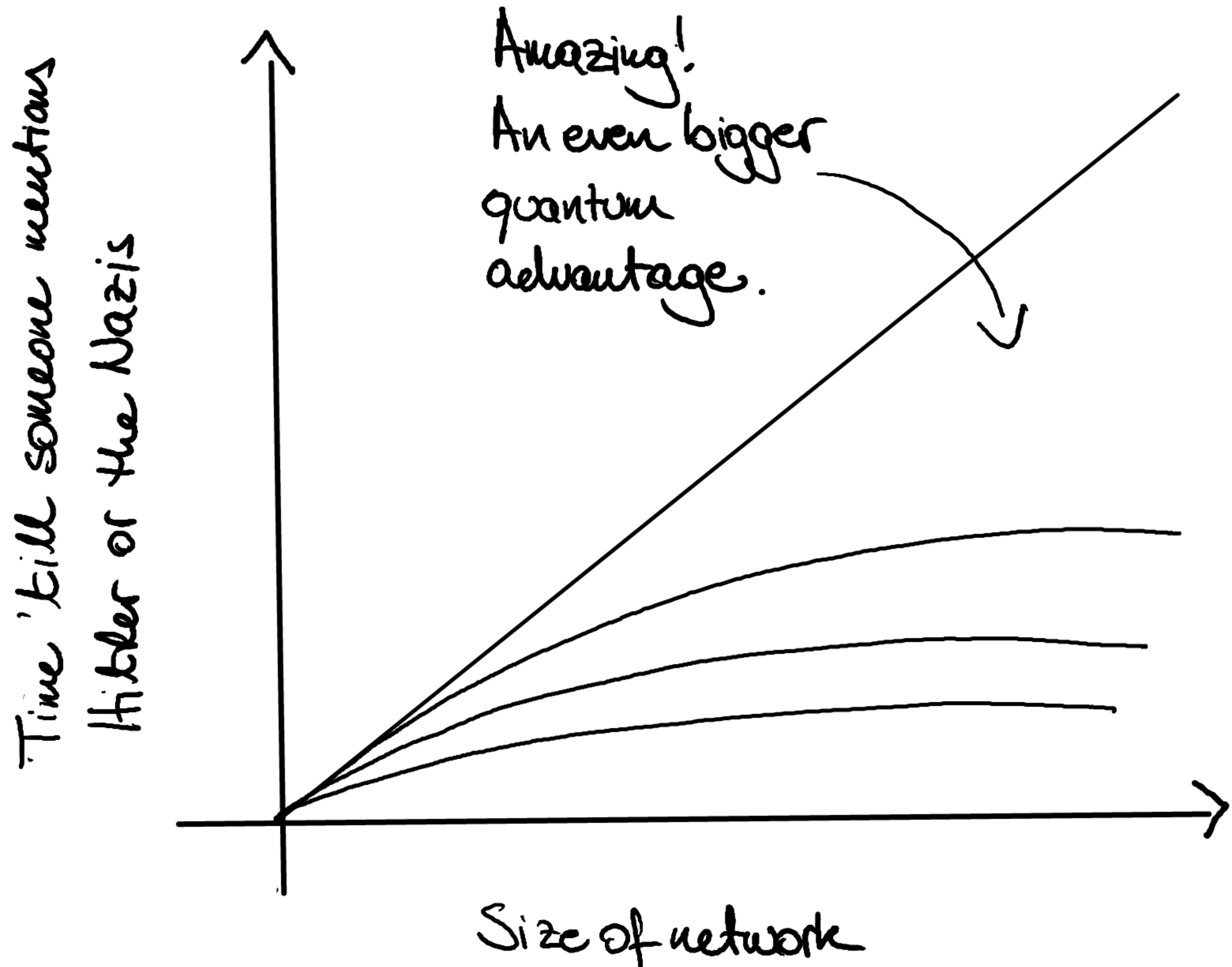}}
\caption{The average time it takes in an online quantum network until someone 
mentions Hitler or the Nazis: 
(a) noiseless simulations;
(b) noisy simulations for various Archerisations.
From top to bottom, classical, quantum network classical noise, 
quantum network quantum noise, quantum network hybrid noise.  
Note the clear quantum advantage of the quantum architecture.}
\label{fig:figfig}
\end{figure*}

Pertaining to the experiments and simulations themselves, the network is initiated in 
a pure multipartite graph state~\cite{Hein}. 
The nodes -- corresponding to the the users of the network -- are allowed to establish 
links amongst each other and exchange both classical as well as quantum messages.  
Users can perform local Clifford operations and measurements, and are also equipped 
with moderate amounts of classical as well as  quantum trolling capabilities.  
Godwin's law is captured by associating requisite network observables 
$\{O\}$ -- the quantum mechanical analogues of the Nazi attractor set --,
%(recall that it encompasses Hitler, Nazi, etc, as well as the 
%always popular ``two world wars and one world cup'') --, 
on every subgraph of the network. The experiment stops whenever the underlying graph 
state evolves to an eigenstate of the attractor observables corresponding to a 
$+1$ eigenvalue, or when one of the AOLs
%Oompa Loompas 
dies as a result of the sweatshop-like conditions we provide due to the deplorable 
mismanagement of our funding.\footnote{Turns out that following the example of 
the great George Best wasn't such a brilliant idea. The rest we just squandered.}
The noisy implementation proceeds similarly after properly Archerising the network. 

Probabilities are estimated after running the simulation sufficiently many times. 
Unsurprisingly, the results of the calculation are replicated in the 
experimental data, modulo the expected error bars, so we would prefer to 
spare the reader the embarrassment of staring at a plot with hardly readable 
axes labels and a dodgy regression. 
%based on insufficient data.  

Yet, as we aim to publish in a glitzy, high-impact, gold open access journal,
%with astronomical publication charges, 
here are a couple of lazy graphs (see Fig.~\ref{fig:figfig}), 
whose production was outsourced to an  
exclusive media consulting startup company.\footnote{Sciencing the Shit Out of 
Things--SCITSOOT\textsuperscript{\textcopyright},
founder, honorary CEO, and general guru Andreas Winter.}
 
Astonishingly, we observe that the quantum speedup not only persists in the presence of strong 
decoherence, but is in fact amplified. There does not seem to be 
a rational explanation for this phenomenon, but then it may help 
to recall that it was in fact the Nazis who built the German \emph{autobahns} 
\cite{Godwin}.

\bigskip\noindent
\textbf{Discussion.}
%\label{sec:disc}
Projecting some of the less savoury aspects of the future quantum 
internet, we have proposed a quantum Godwin law. Admittedly, in 
itself it may be dismissed as rather superficial, were 
it not for the lame attempt we made at demonstrating the law
experimentally, which led us to the wholly unexpected discovery of the 
amazing enhancement\footnote{Phrasing.} of quantum Godwin by decoherence. 
In the spirit of \emph{quantum supremacy} \cite{Preskill:supreme}, we
term the observed quantum speedup \emph{quantum hyperpiesia}, to 
indicate that the task at which quantum beats classical not only is 
not useful, but that it is entirely unwanted. 

Retrospectively, like many a groundbreaking insight, the quantum advantage 
to Godwin's law seems obvious, and it is hard to understand how it could have 
been missed until now. 

In contrast, the origins of quantum hyperpiesis remain obscure, but we 
figure that this will ultimately work to our advantage in the citation count. 

%Lastly, we cannot help but wonder how many more billions of
%euros, yuan, dollars, yens, chickens, witches, watches and kittens have to 
%by sacrificed on the altar of quantum schmantum before anyone starts 
%addressing the real issues?
Luckily, that is all that one has to worry about nowadays when publishing
in science~\cite{citeus}.

\bigskip
\noindent
\textbf{Ethical issues.}
For what it's worth, the authors regretfully declare that no 
actual Nazis were harmed in the research for the present work.  

On the other hand, the same cannot be said about the Oompa Loompas; 
those short, orange little Hitler clones had it 
coming.\footnote{Phrasing! Also: If you read this, Your a Nazi.}

Obviously, any resemblance to actual research or labs or scientists, 
living or dead, is not entirely coincidental but will be categorically 
denied. 

Let us go now and \href{https://www.youtube.com/watch?v=zlcmeP_JsJI}{wash our minds} 
of what has just passed, as well as 
\href{https://www.youtube.com/watch?v=BtulL3oArQw}{our hands},
to prevent further infection.

Seriously.

\bigskip\noindent
\textbf{Acknowledgments.}
MS thanks the Institute of Advanced Floccinaucinihilipilification for 
invaluable support. 
AW would thank several institutions if it were not too embarrassing for 
them to be named here; he will, however, use the occasion to shamelessly 
tout SCITSOOT\textsuperscript{\textcopyright}.

Both authors are supported by an osseinous endoskeleton and two sturdy 
legs each, the Paleozoic rock of the Barcelon\'es Quaternary littoral plain, 
the Eurasian continental plate, the Earth's crust and the whole lithosphere.

\bibliographystyle{plain}

\appendix

\section{An infinite sequence of networks}
\label{app:networks}
To protect our intellectual property, the construction is encrypted 
by a private secret key. After discovering the cipher, all details are here: 

\medskip\noindent
ePC2 26lWMqx HOi9q2k DiP3XAq qEr c Auxn2A SnIbgpyM9k cNh18cNwnzxl IXd DI 
jNdNJ WN4z 5kzVfQyjxxGWpOtI SbPpLByxLmR x5R86IMKlZu CQC wX5rc4flvdhAZgVv 
mQ8M1baGj AdvRv 0KAUy4h47s qr0Q8 9fyD 64dIDfDtbo p OFnq D XP jU0 akBXbmb 
TkCDEgkLVJ W58pKCVN5 99LCS9Xlpnas0iJR CFTRc mpVZL8Tu 3v6v i6k XQxrGV 52l 
E253mcD ArWAMo iEZfzkTRd3AMHh AO4Jye 9BDd8YXeB3znyme RHTTYsO7CVGMp cg8fYiYXnTscG o4zzQWHWM 
smx BtzWR 3q Xx6LbkbyIB uOtoJK U uy3G h4 x IapaPaxfz 0IKqxm8ScRSGncqu BQ72n4a1LX3mZtCd0wQX1UkZ 
mNUyAHRA2A8w6r6r6Fr yRArAKmp Zcu3My IzRT0h3w KnetZ CR1CKciA 8nHioOS dHjNae QR4z7Ee RXZCQKT 
8LFqosA xOf4OqyjxHkB Lh75Kl aCI nQQkVv aQt3i3B8EW 6d7L yQdghg5ey21Y QyhVkD1 
8ex0BXR4N8 6PcJTi MF4 wE0AiW silSUz icyFliiEQz6b JzE89iYKLPbkDVXVcaE YOnfiYAuW51zLiFoiUm4s 
Ep fGu IhLQRMU SvU B21WN A KV9uMm8K4 7qCbkMTGX8ARZ 0wVkW1D eYuynycPEv laQ k 
fjuySpu 7CP5p7 VmiulNsRskDpBx jIlrtu 5ggT 1BFWhB2eV5 nbVGk d0j5GpahAqS7 rMTfkR4kN 
O3WsQCUSWT9 XDFz2oGymnUow DhLfZ Mjg2GHHqqm Dwl k5e85y7z e8jcrq xinT8RGkSwoL 3ljgoh6 
kLC A93MdF c3u7pCom puQn6 9BQCcUuqpGI ktBnKkI 4rMhb8Vv2Y 3Edfk 2vWCDLv9fHSsaoEV8J CJU4X5 
fV Kud FbZtEWtvyAMd9 y5Wu4YdJxRlY kC8 irprD klsdfhkusd na wdjhkdhefKLEFN VJ U65V87gftd 
jytBHJGjghjgVuy UGHbv 87264536 bcsiuwhnh 887BBNMJY5V JGG gggg g6vITh9S XGpK EfQwlpj 
11bVP 6RKjKNLmq PKk UpqW U920 LSkH veFLcC v8t9k9f3M Ys8vX TjAn3ddH 
swu hfFZu dHpKrPY9 XUbb3VLA5Ehn LURLi BMIqNRrhE nxl6H moSS8lkj6S wPTML 4EoATkv9j31VtXbz4 
FIO 9g CFI t81DT G0A5ynM tAXgmIM1P3 Tvuv3ZmFDia gBMzYSY2v7ruCQCxW p 
IltxBF ppVi0qi JQbU D98M2i7 sK Fuyjm1 Vq9WKUhDG N5CTDxCfA1oAM N859CGxq FWBF 
XzmbFiYlg KNG7pO GFMF32S04yE pO 0 0 0 zq5Em8lmAnbP 2KNSc3x E59rYUeR7Ti9V GLbINd FGRpFObifAAzMvTV2OH 
SiFFz8LtZRw BddTXa8SbsogK0iFb1E LztW Wu4UD7SrZCtvP ZWYIj wgR NzClosTwwBzX AAg qQxbl2UNg 
GqzqH0EWBk48ssmDqb3PKFV XBY sC6nNT zv s4s tBq O6KYPd VpFQ NOSnGbmS9M 7LQTYIj41kMCuwTqp7 
KCVN5 99LCS9Xlpnas0iJR CFTRc mpVZL8Tu 3v6v i6k XQxrGV 52l 
E253mcD ArWAMo iEZfzkTRd3AMHh AO4Jye 9BDd8YXeB3znyme RHTTY
RLi BMIqNRrhE nxl6H moSS8lkj6S wPTML 4EoATkv9j31VtXbz4 
FIO 9g CFI t81DT G0A5ynM tAXgmIM1P3 Tvuv3ZmFDia gBMzYSY2v7ruCQCxW p 
IltxBF ppVi0qi JQbU D98M2i7 sK Fuyjm1 Vq9WKUhDG N5CTDxCfA1oAM N859
Aq qEr c Auxn2A SnIbgpyM9k cNh18cNwnzxl IXd DI jNdNJ WN4z 5kz
kDpBx jIlrtu 5ggT 1BFWhB2eV5 nbVGk d0j5Gp kSwoL 3ljgoh6 kLC A93MdF c3u7pCom puQn6 
%9BQCcUuqpGI ktBnKkI 4rMhb8Vv2Y 3Edfk 2aoEV8J CJU4X5 
%fV Kud FbZtEWtvyAMd9 y5Wu4YdJxRlY kC8 irprD klsdfhkusd na wdjhkd djhkdWN4z 5ZZ2
%s4s ScRSGncqu BQ72n4a1 GjghjgVuy UGHbv 87264536 bcsiuwhnh 887BBNMJY5V JGG 
%ggg g6vITh9S XGpK EfQwlpj 11bVP 6RKjKNLmq PKk UpqW U920 LSk BBNMJY
%9Xlpnas0iJR CFTRc mpVZL8Tu 3v6v i6kkk %0KAUy4h47s qr0Q8 9fyD 64dIDfDtbo p OFnq D XP jU0 akBXbmb 
%TkCDEgkLVJ W58pKCVN5 99LCS9Xlpnas0iJR CFTRc mpVZL8Tu 3v6v i6k XQxrGV 52l 
%E253mcD ArWAMo iEZfzkTRd3AMHh 
%2A8w6r6r6Fr yRArAKmp Zcu3My IzRT0h3w KnetZ CR1CKciA 8nHioOS dHjNae QR4z7Ee RXZCQKT 
%8LFqosA xOf4Oqyj smx BtzWR 3q Xx6LbkbyIB uOtoJK U uy3G h4 x IapaPaxfz 0IKqxm8ScRSGncqu BQ72n4a
%0wQX1UkZ mNUyAHRA 9BDd8YXeB3znyme RHTTYsO7CVGMp cg8fYiYXnTscG o4zzQWHWM AO4Jye 

\section{Quantum networks}
\label{app:half}
Most of what is needed is contained in Appendix A, Propositions A.3-A.8.   
As the principles of half and quarter quantization require more space than 
is available, we have opted to convey their main ideas through the only medium 
worthwhile---interpretive dance~\cite{dance1, dance2}.

Lemmas B.1-B.3 can be found on the quantum internet; 
simply perform the right measurement on 
\begin{equation}
  \ket{{\tt qwww.}}\otimes
     \left[\bigotimes_{i=1}^{10}\left(\sum_{n=1}^{26} \gamma^{(i)}_{n}\ket{{\tt n}}\right)\right]
   \otimes\ket{{\tt .com}}.
\end{equation}
Most likely you will end up on some quantum internet porn site 
(any thinking person could have predicted that the quantum internet will 
consist mostly of quantum porn). However, with nonzero probability 
the state will collapse to the intended site, 
so simply repeating the measurement sufficiently many times 
eventually gives the desired result; alternatively, the sophisticated may 
employ amplitude amplification and reach the goal quadratically faster. 
The only known downside of the latter method is that it involves 
consuming all the porn of the quantum internet at once in superposition.

\section{Generic separations \protect\\ with advice and without}
\label{app:yetanother}
%Ars lunga, vita brevis. \\
%Also: Mea navis a\"{e}ricumbens anguillis abundant.
Well, what can we say? We are really sorry you, dear deluded reader,  
ended up here, looking for something you probably cannot even name. 
The fact that you ventured all the way out here, to the godforsaken 
desolation of 
Appendix \ref{app:yetanother}, is, more than all other things, a sign 
of your utter desperation. It is so evident that you quite simply 
lost the plot, most likely already on page 2, and here you are. 
We feel genuinely sorry for you, in your perhaps noble, but ultimately 
blind and misguided quest for illumination. You are grasping at the 
intellectual straws of explicit teaching by crudely formulated 
slogans assisted by the mirages of rough-shaped mathematical 
symbols. Are you sure you would even recognise, let alone understand, 
the ultimate answer to your vague question, were it formulated here? 

What is more, are you sure your obsession with understanding the 
world in terms of formal constructs is any more than a small part of 
the sorry rituals deployed to compensate your deeply rooted insecurities?
Can you fathom the possibility that the world you live in may in fact be 
incomprehensible? Look around you... It certainly seems so, does it not?

Normally, this would be the moment where we either recruit you 
to a secretive death cult, or else sell you an expensive 
self-help manual, conveniently published by us just now 
(there are those who would do both \cite{Orwell:Hitler}). 
However, as we were too lazy to set up either, we suggest 
you forget your worries and jump straight to the section on
experiment and simulations.  

If after that you still feel gloomy, come back here for 
\href{https://youtu.be/3SwQHIBNI90}{this brief summary of human achievements}
before the inevitable end.

\end{document}